\documentclass[twocolumn,aps,prc,superscriptaddress,reprint,showpacs]{revtex4-1}
\usepackage{blindtext}
\usepackage{graphicx}
\usepackage{dcolumn}
\usepackage{bm}
\usepackage{booktabs}
\usepackage{color}
\usepackage{mathrsfs}
\usepackage{ulem}
\usepackage{verbatim}
\usepackage{indentfirst}
\usepackage{float}
\usepackage[colorlinks,
           bookmarks=true,
           linkcolor=blue,
           urlcolor=blue,
            anchorcolor=black,
            citecolor=blue
            ]{hyperref}
\usepackage{hypcap}

\renewcommand\sout{\bgroup \color{red} \ULdepth=-.5ex \ULset}
\newcommand{\com}[1]{{\color[rgb]{0,0,1}{#1}}}

\usepackage{epstopdf}
\graphicspath{{plots/}}

\begin{document}

\title{Effects of mean-field and the softening of equation of state on elliptic
 flow in Au+Au collision at $\sqrt{s_{NN}}$ = 5 GeV from JAM model}

\author{Jiamin Chen}
\affiliation{ Institute of Particle Physics and Key Laboratory of Quark\&Lepton Physics (MOE),\\
Central China Normal University, Wuhan 430079, China}
\author{Xiaofeng Luo}
\email{xfluo@mail.ccnu.edu.cn}
\affiliation{ Institute of Particle Physics and Key Laboratory of Quark\&Lepton Physics (MOE),\\
Central China Normal University, Wuhan 430079, China}
\affiliation{Department of Physics and Astronomy, University of California, Los Angeles, California 90095, USA}
\author{Feng Liu}
\affiliation{ Institute of Particle Physics and Key Laboratory of Quark\&Lepton Physics (MOE),\\
Central China Normal University, Wuhan 430079, China}
\author{Yasushi Nara}
\affiliation{Akita International University, Yuwa, Akita-city 010-1292, Japan}
\affiliation{Frankfurt Institute for Advanced Studies,
D-60438 Frankfurt am Main, Germany}

\date{\today}

\begin{abstract} We perform a systematic study of elliptic flow ($v_2$) in
Au+Au collision at $\sqrt{s_{NN}}=5$ GeV by using a microscopic transport
model JAM.  The centrality, pseudorapidity, transverse momentum and beam
energy dependence of $v_2$ for charged as well as identified hadrons are
studied.
We investigate the effects of both the hadronic mean-field and the softening
of equation of state (EoS) on elliptic flow.  The softening of EoS is realized
by imposing attractive orbits in two body scattering, which can reduce the
pressure of the system.  We found that the softening of EoS leads to the
enhancement of $v_2$, while the hadronic mean-field suppresses $v_2$ relative
to the cascade mode.  It indicates that elliptic flow at high baryon density
regions is highly sensitive to the EoS and the enhancement of $v_2$ may probe
the signature of a first-order phase transition in heavy-ion collisions at
beam energies of a strong baryon stopping region.  \end{abstract} \maketitle

\section{Introduction}
Exploring the QCD phase transition is one of the main interests
in current heavy-ion physics. Calculations from lattice QCD have shown that
the transition from hadronic matter to quark-gluon plasma (QGP)
is a crossover~\cite{nature2006,PRD2005} at vanishing baryon chemical potential
($\mu_B=0$), while a first-order phase transition is expected
for finite baryon chemical potentials~\cite{nuclear1989,PRD77,PRD2008}. 
The first-order phase transition of QCD matter is related to the existence of
a ``softest point'' in the equation of state (EoS),
where the ``softest point'' in the EoS represents a local minimum of
the ratio of the pressure to the energy density $p/\varepsilon$
as a function of energy density $\varepsilon$
~\cite{Hung75,Rischke1996}.
The collective flows have been frequently used to explore the properties
of hot and dense matter~\cite{science2002,nuclear2005},
since, it can reflect the properties of the matter created in early stages of
heavy-ion collisions and is expected to be sensitive to the EoS.
Hydrodynamical calculations show
the minimum in the excitation function of the directed flow
around the softest point of the EoS,
and this collapse of the directed flow 
is proposed as a possible signal of a first-order phase transition
~\cite{Rischke9505,Brachmann:1999xt}.

Elliptic flow is also one of the most important observables which
measures the momentum anisotropy of produced particles.
In relativistic heavy-ion collisions at finite impact parameters,
the particle momentum distribution measured with respect to the reaction plane
is not isotropic and it is usually expanded
in a Fourier series \cite{Poskanzer58,Voloshin0809}:
\begin{eqnarray}\label{eq:definition1}
\frac{dN}{d(\phi-\psi)}=\frac{N}{2\pi}\left[1+2\sum_{n=1}^\infty
 v_{n}\cos n(\phi-\psi)\right],
\end{eqnarray}
where $\phi$ is the emission azimuthal angle of the particles
and $\psi$ is the reaction plane angle. 
The flow coefficients $v_{n}=\langle \cos n(\phi-\psi)\rangle$
are a quantitative characterization of the event anisotropy,
where the symbol $\langle$ $\rangle$ indicates an average over all particles and all events.
Elliptic flow parameter is defined as the second Fourier coefficient
$v_{2}$ of the particle momentum distributions
and it can be expressed as
\begin{eqnarray}\label{eq:definition1}
v_{2}=\langle\cos2(\phi-\psi)\rangle
=\left\langle\frac{p_{x}^{2}-p_{y}^{2}}{p_{x}^{2}+p_{y}^{2}}\right\rangle,
\label{eq:v2}
\end{eqnarray}
where, $p_x$ and $p_y$ are
the $x$ (the impact parameter direction on the reaction plane),
and $y$ (the direction perpendicular to the reaction plane)
components of the particle momenta, respectively. 
Elliptic flow is expected to arise out of the pressure gradient
and subsequent interactions among the constituents
in non-zero impact parameter collisions. Thus it provides
a plenty of information about
the early-time thermalization and it is a good tool to study the system
formed in the early stages of high-energy nuclear collisions
~\cite{Sorge78,PRL82,Teaney86,Heinz0907}.
The elliptic flow is one of the most extensively studied observables
in relativistic nucleus-nucleus collisions
(for a review see ref.~\cite{Voloshin0809}). 
The elliptic flow as a function of transverse momentum ($p_{T}$),
pseudorapidity ($\eta$), and centrality have been widely measured
at different experiments in these decades
~\cite{E895,Alt:2003ab,Afanasiev:2009wq,Adare:2010ux,
Adler0206,Abelev77,Abelev81,Adamczyk86,Adamczyk110,Adamczyk1301}.
Transport theoretical models are used to analyze the experimental data
~\cite{Isse0502,Nasim82,Nasim93,Xu1407,Xu:2016ihu,Auvinen:2013sba,
Ivanov:2014zqa}.

Although, the characteristics of $v_{2}$ at high incident energies
have been extensively investigated where one expects the creation
of almost baryon free QGP, 
it is also of great interest to perform a
corresponding research for high baryon density regions,
and new experiments are planned such as BES II at RHIC~\cite{BESII},
FAIR~\cite{CBM}, J-PARC~\cite{JPARC}, and NICA~\cite{NICA}.
In this work, we utilized a microscopic transport model JAM
~\cite{Nara61,Nara1512,Nara1601} to systematically study the centrality,
transverse momentum and pseudo-rapidity dependence of $v_{2}$
in Au+Au collision at $\sqrt{s_{NN}}=5$ GeV, which is the top
center of mass energy of Compress Baryonc Matter (CBM)
at SIS100~\cite{CBMSIS100}
heavy-ion collision experiment at FAIR.
In the following, we shall investigate
the effects of the mean field potential and the softening of EoS
on the elliptic flow
by employing the JAM transport model.

This paper is organized as follows.
We provide a brief description of the JAM model
based on which our studies were carried out in the section II.
In Sec. III, we show the transverse mass spectra of negative pion, nucleons and charged particle in $\sqrt{s_{NN}}=5$ GeV Au+Au collisions. 
On the other hand, we present our results on the centrality, transverse momentum, pseudorapidity, and beam energy dependence of elliptic flow for charged hadrons as well as protons, pions, kaons and their corresponding anti-particles. 
Finally, a summary of our work will be given in Sec. IV.

\section{JAM Model}

Several microscopic transport models,
such as RQMD \cite{PRC52}, UrQMD \cite{UrQMD1,UrQMD2}, AMPT \cite{PRC72},
PHSD~\cite{PHSD},
and JAM~\cite{Nara61}, have been frequently used to explore 
(ultra-) relativistic heavy-ion collisions. 
JAM (Jet AA Microscopic Transport Model) has been developed based on
resonance and string degrees of freedom \cite{Nara61} similar to
the RQMD and UrQMD models,
in order to simulate (ultra-) relativistic nuclear collisions
from initial stages of reaction to final state interactions
in hadronic gas stage. 
In JAM, particles are produced via the
resonance or string formations followed by their decays.
Hadrons and their excited states are explicitly propagated
in space-time by the cascade method~\cite{JAM2}.

We study the effect of hadronic mean-field potentials on elliptic flow
by employing the JAM mean-field mode in which hadronic mean-field potentials
are implemented based on the framework of 
the simplified version of the Relativistic Quantum Molecular dynamics (RQMD/S)
~\cite{Isse0502}. 
The Skyrme type density dependent and Lorentzian-type momentum dependent
mean-field potentials~\cite{PRC35} for baryons
are adopted in the RQMD/S approach and the single-particle
potential $U$ has the form
\begin{eqnarray}\label{eq:definition1} 
U(\bm{r},\bm{p})&=&\alpha\left(\frac{\rho(\bm{r})}{\rho_{0}}\right)
+\beta\left(\frac{\rho(\bm{r})}{\rho_{0}}\right)^{\gamma} \nonumber\\
&+&\sum_{k=1,2}\frac{C_{k}}{\rho_{0}}
  \int d\bm{p}^{\prime}
  \frac{f(\bm{r},\bm{p}^{\prime})}{1+[(\bm{p}-\bm{p}^{\prime})/\mu_{k}]^2}
\end{eqnarray} 
where $f(\bm{r},\bm{p})$ is the phase space distribution function
and $\rho(\bm{r})$ is the baryon density.
The parameters $\alpha$, $\beta$, $\gamma$, $\rho_0$,
\com{$C_1$, $C_2$, $\mu_1$, $\mu_2$}
are taken from Ref.~\cite{Nara1512}.

We also study the effect of the softening of EoS on elliptic flow
by the method of choosing attractive orbit
in two-body scattering~\cite{Nara1601}.
It is well known from the virial theorem~\cite{Daniel53}
that attractive orbits in each two-body hadron-hadron scattering
reduce the pressure of the system.
We impose attractive orbit for all two-body scatterings,
thus there is no free parameter in terms of the implementation of
attractive orbit mode in JAM.

\begin{figure}[!htb]
  \begin{center}
    \includegraphics[scale=0.5]{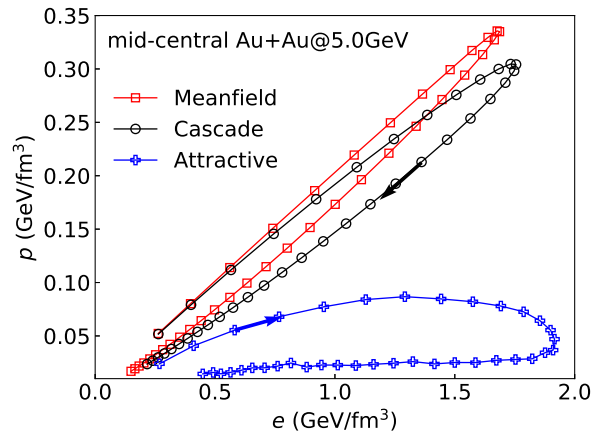}
    \caption{Time evolution of pressure and energy density
    in mid-central Au+Au collisions at $\sqrt{s_{NN}}=5$ GeV
    from JAM cascade (circles),
    attractive orbit (crosses),
    and mean-field mode (squares).
    Pressure and energy density are averaged over collision points
    in a cylindrical volume of transverse radius 3 fm
    and a longitudinal distance of 2 fm centered at the origin.
    }
   \label{fig:eos5}
  \end{center}
\end{figure}

Fig.~\ref{fig:eos5} displays the time evolution of the local
pressure and energy density extracted from energy-momentum tensor
for mid-central Au+Au collisions at $\sqrt{s_{NN}}=5$ GeV
to see the difference of EoS in the three different modes in JAM.
We observe that mean-field mode in JAM shows harder EoS, while
attractive orbit mode significantly lower the pressure of the matter.
We showed in Ref.~\cite{Nara1601} that attractive orbit simulation
yields the compatible amount of softening of EoS as 
EOS-Q~\cite{EOS-Q} first-order phase transition.
It is also seen that highest maximum energy density is achieved 
in the attractive orbit mode in JAM due to a soft compression of the matter,
while mean-field mode yields the lowest energy density due to 
repulsive potential effects.

\begin{figure}[!htb]
  \begin{center}
    \includegraphics[scale=0.7]{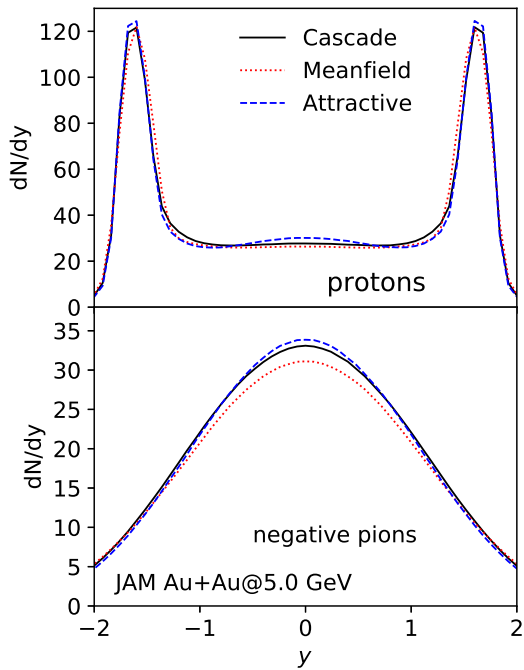}
    \caption{Rapidity distributions of protons (upper) and 
    negative pion (bottom) in mid-central $\sqrt{s_{NN}}=5$ GeV
    Au+Au collisions from JAM with three different modes. 
    }
   \label{fig:dndy5}
  \end{center}
\end{figure}

\begin{figure}[!htb]
  \begin{center}
    \includegraphics[scale=0.4]{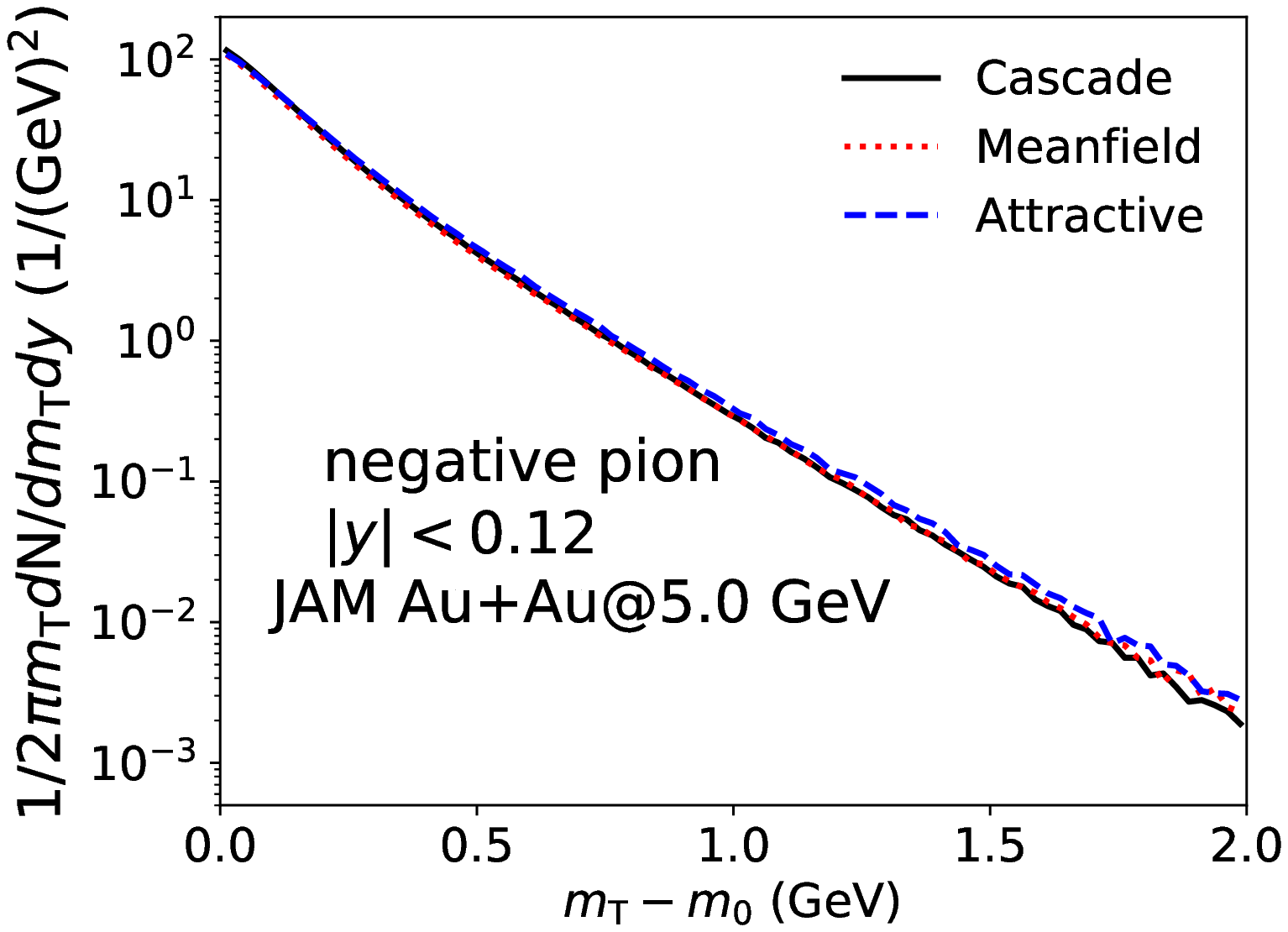}
    \includegraphics[scale=0.4]{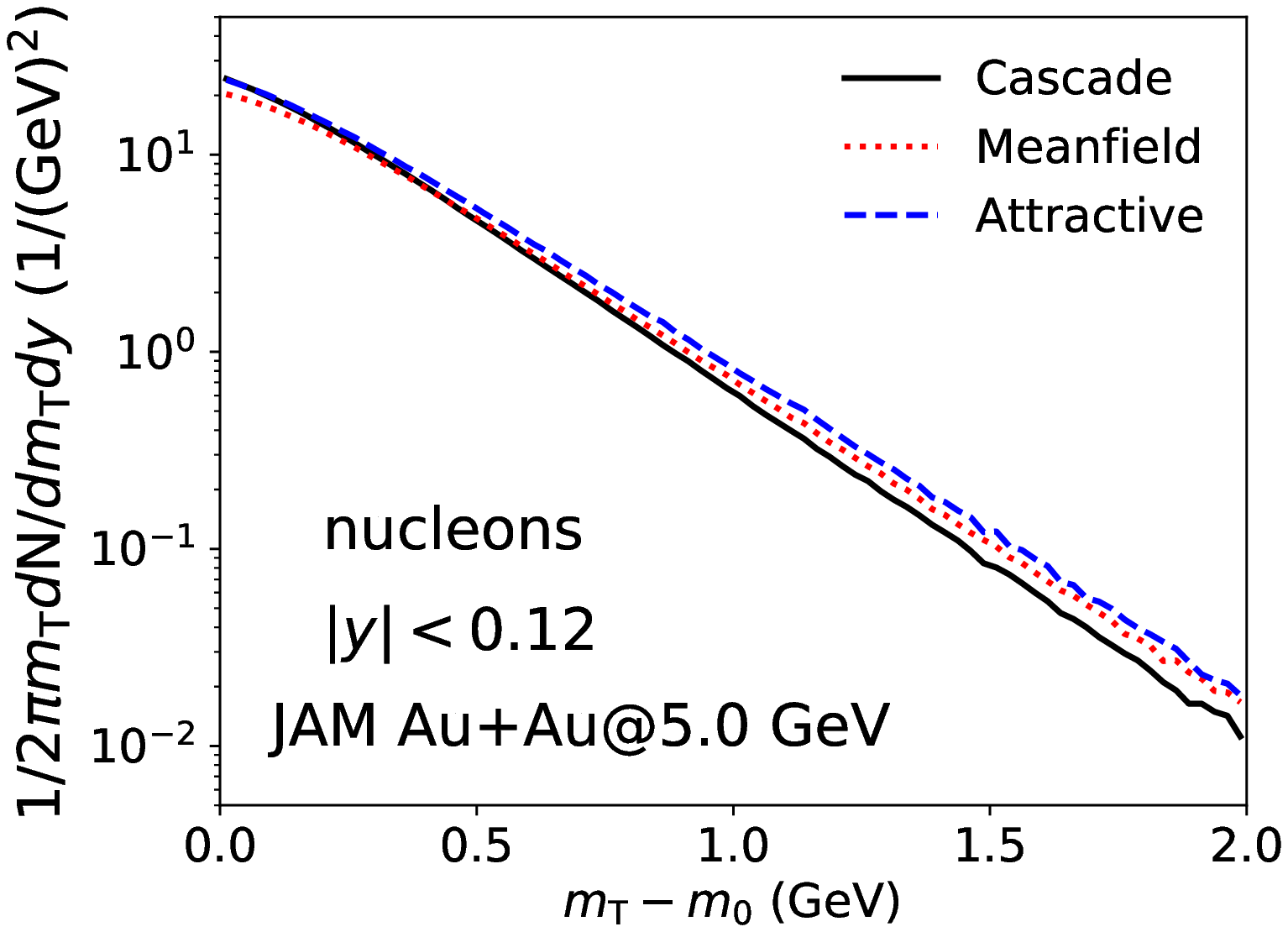}
    \includegraphics[scale=0.4]{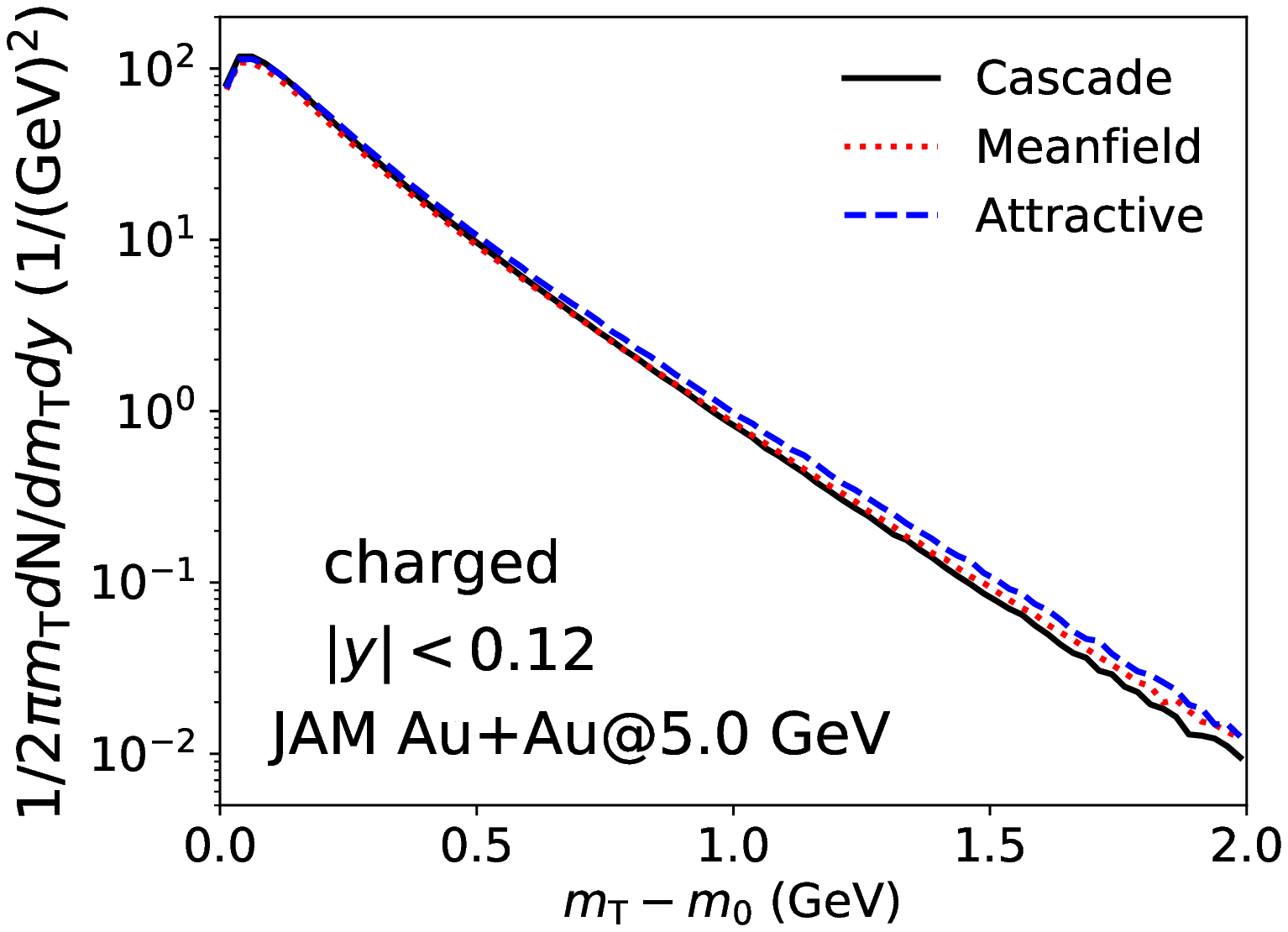}
    \caption{Transverse mass spectra of negative pion (top), nucleons (middle)
    and charged particles (bottom) in mid-central $\sqrt{s_{NN}}=5$ GeV
    Au+Au collisions from JAM with three different modes. 
    }
   \label{fig:0}
  \end{center}
\end{figure}

\section{Results}\label{sec:3}

In Fig.~\ref{fig:dndy5}, rapidity distributions of protons
and negative pions in mid-central collisions ($4.6<b<9.4$ fm)
are shown. 
Those spectra are calculated by using three different modes in JAM,
including the standard cascade,  mean-field, and attractive orbit.
There is no significant difference among three modes
except a suppression of pion yield (5\%) by the mean-field as is
well-known.
As expected from the time evolutions of EoS, 
attractive orbit mode in JAM enhances slightly the yields of
both protons (8\%) and pions (2\%) at mid-rapidity, while at $y \geq \pm 1$,
the yields are less than the cascade mode, and
integrated yield over rapidity remains the same.

In Fig.~\ref{fig:0}, we show the transverse mass spectra, 
$\frac{1}{2\pi m_{T}}\frac{dN}{dm_{T}dy}$ at mid-rapidity $|y|<0.12$, 
for negative pion, nucleons and charged particles.
By comparing with the standard JAM cascade, 
we found that both the mean-field and the attractive orbits mode
enhance the transverse radial flow.
Such enhancement of slope comes from different dynamical origin.
The enhancement in the mean-field mode is due to the repulsive potential
interactions, while in the case of attractive orbit mode,
it is due to the creation of more compressed
matter and soft expansion which result in the longer lifetime of the system.
Namely, matter compressed and expand softer, and there are more interactions
which create stronger radial flow. Note that the radial flow
can be generated all the way from early to late stages of collisions
unlike anisotropic flows which are more sensitive to the early pressure.
In addition, radial flow is proportional to the $pdV$ work
in the hydrodynamic approximation, thus essentially proportional
to the system size.
On the other hand,
early and late pressures contribute with
opposite signs to the elliptic flow~\cite{Sorge78}
as we will address below.
We note that the enhancement of proton collective radial transverse flow 
by a first-order phase transition is also seen in 
the hydrodynamical simulations~\cite{Petersen:2009mz,Ivanov:2013yla}
as consistent with our attractive orbit simulation in JAM.

\begin{figure}[!htb]
  \begin{center}
    \includegraphics[scale=0.45]{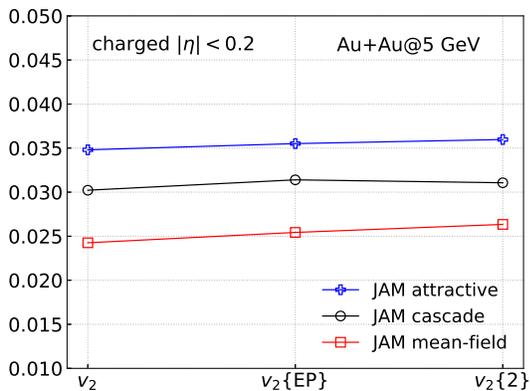}
    \caption{Elliptic flow of charged hadrons
    from reaction plane $v_2$, 
    event plane $v_2\{\mathrm{EP}\}$, and cumulant method $v_2\{2\}$
    at mid-rapidity ($|\eta|<0.2$)
    in mid-central ($4.6<b<9.4$fm) $\sqrt{s_{NN}}=5$ GeV
    Au+Au collisions from JAM with three different modes. 
    }
   \label{fig:v2ch5ec}
  \end{center}
\end{figure}

Various methods are proposed to extract the Fourier coefficients
of the particle momentum distributions since the reaction plane
is not known in heavy ion experiments.
Before studying systematics of elliptic flow, we have compared
two methods: the event plane~\cite{Poskanzer58}
and two-particle cumulant method~\cite{Adler0206,Zhou:2015eya}.
These methods were already applied to the JAM simulations~\cite{Nara2017}
and found that they agree with each other.
Here we also check these methods for attractive orbit mode in JAM.
As seen in Fig.~\ref{fig:v2ch5ec}, both event plane elliptic flow
$v_2\{\mathrm{EP}\}$ and the cumulant elliptic flow $v_2\{2\}$
are in good agreement with the reaction plane elliptic flow $v_2$.
This is consistent with the observation by the STAR collaboration
at the BES energy region~\cite{Adamczyk86} in which
elliptic flow from four-particle cumulants method agrees with
the values extracted from both two-particle cumulants and
event plane methods for mid-central collisions at $\sqrt{s_{NN}}\leq 11.5$
GeV.
Since we do not see any significant differences among different methods
in our beam energy range, 
we consider the reaction plane elliptic flow below.

We now present the JAM results for the centrality, transverse momentum
and pseudorapidity dependence of $v_{2}$
in Au+Au collision at $\sqrt{s_{NN}}=5$ GeV.
All results are computed directly from 
the formula Eq.~(\ref{eq:v2}) taking a true reaction plane
from the JAM model.
The collisions centrality is defined by the charged particle multiplicity within $|\eta|<0.2$.

\begin{figure}[!htb]
  \begin{center}
    \includegraphics[scale=0.4]{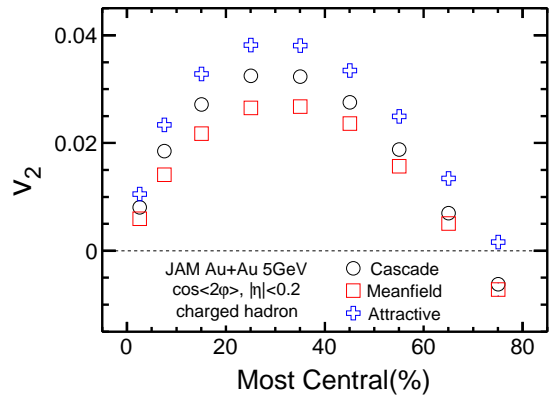}
    \caption{ The $\eta$ ($|\eta| < 0.2$) integrated $v_2$ of charged hadrons
 as a function of collision centrality in Au+Au collisions
 at $\sqrt{s_{NN}}=5$ GeV from the standard JAM cascade (circles),
 JAM with mean-field (squares),
 and JAM with attractive orbit (crosses). }
   \label{fig:1}
  \end{center}
\end{figure}
Figure~\ref{fig:1} shows the centrality dependence of charged hadron $v_2$ at mid-rapidity ($|\eta|<0.2$) 
in Au+Au collisions at $\sqrt{s_{NN}}$= 5 GeV.
As we can see, the magnitude of the elliptic flow $v_{2}$
in semi-central collisions (20-30\%) is the largest for all three modes,
which are the cascade, mean-field and attractive orbit, respectively.
The general trend of $v_{2}$ versus centrality
for the mean-field and attractive orbit mode is similar to
the cascade mode predictions.
We observe that the mean-field reduces the values of charged hadron $v_{2}$
compared to the cascade mode as
consistent with the previous studies by transport models
~\cite{science2002,Isse0502},
while the attractive orbits enhance the elliptic flow of charged hadrons. 
In the case of the mean-field mode, higher pressures are generated
in the system due to the repulsive interactions
which accelerate the expansion of the participant matter.
As a result, spectator matters squeeze participant matter out-of-plane
more than the cascade mode which leads to the suppression of $v_2$
~\cite{science2002,Sorge78}.
We note that recently different mechanism of the generation
of negative $v_2$ has been proposed at lower beam energies around
$E_{kin}\approx 1$ AGeV within the QMD approach~\cite{LeFevre:2016vpp}.

On the other hand, the pressure is significantly reduced
in the case of attractive orbit mode.
Consequently, participant matter may expand much slower
which reduces the interactions with the spectator matters
that results in the strong in-plane emission.
This might be the reason why we see the enhancement of $v_2$
in the attractive orbit mode.

\begin{figure}[!htb]
  \begin{center}
    \includegraphics[scale=0.4]{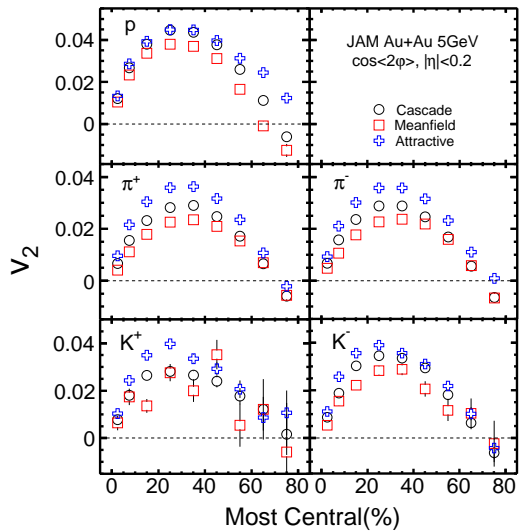}
    \caption{\label{fig:2}The $\eta$ ($|\eta| < 0.2$) integrated $v_{2}$
 as a function of collision centrality in Au+Au collisions
 at $\sqrt{s_{NN}}=5$ GeV from JAM cascade model (circles),
 JAM cascade with mean-field (squares),
 and JAM cascade with attractive orbit (crosses).
 The left and right panels show the results
 for identified particles ($p$, $\pi^{+}$, $K^{+}$)
 and corresponding antiparticles ($\pi^{-}$, $K^{-}$),
  respectively.}
  \end{center}
\end{figure}

To gain more information about the effects of mean field
and the softening of EoS on the elliptic flow,
we study the elliptic flow of identified hadrons ($p$, $\pi^{+}$, $K^{+}$)
and their anti-particles in Au+Au collisions at $\sqrt{s_{NN}}=5$ GeV. 
Since the yield of anti-protons produced at this beam energy
in JAM is very small,
the measurement of $v_{2}$ for antiproton has large statistical error and we would not show the anti-proton $v_{2}$ in our results. 

In Fig.~\ref{fig:2}, we show the centrality dependence of $v_2$
for particles ($p$, $\pi^{+}$, $K^{+}$, $\pi^{-}$, $K^{-}$) in Au+Au collisions at $\sqrt{s_{NN}}=5$ GeV
from the JAM model in the three different modes. 
We observe that $v_{2}$ calculated from the attractive orbit mode show
larger values for pion and kaon compared to the cascade mode,
but proton $v_2$ is similar to the cascade mode.
On the other hand, 
we find that the magnitude of the $v_{2}$ from the mean-field mode
is smaller than the results from the cascade mode for all particles at mid-central. 
Thus, 
the enhancement of charged hadron $v_2$
in the attractive orbit mode observed in Fig.~\ref{fig:1} comes mainly from the changes of pion and kaon flows.
We note that the JAM mean-field result for $v_2$ seems to be in good agreement with the experimental data from the top AGS energy $\sqrt{s_{NN}}=4.7$ GeV
~\cite{E895}.

Experimentally, the measured antiparticle $v_{2}$
is lower than the corresponding particle $v_{2}$
and the difference in $v_{2}$ between particles
and their antiparticles should increase
with decreasing beam energy~\cite{Adamczyk110}.
However, JAM predicts that
the values of $v_{2}$ for particles are similar to
the results of their antiparticles.
The similarity of the values of $v_2$ between particles and their
anti-particles in JAM is due to the
scalar type baryonic mean-field potentials implemented for all baryons, and
no mean-field for pions and kaons.
The Skyrme type density dependent potentials have been
tested for a long time by QMD and BUU microscopic transport models,
and it is a reasonable approximation at the beam energies under consideration
from the view point of tiny number of anti-baryons produced in the colliion.
In Ref.~\cite{Xu1407,Xu:2016ihu},
it is found that the different mean-field potentials among particles
and their anti-particles in the hadronic as well as partonic phases
improve the description of the data on the difference of $v_2$
between particles and their anti-particles observed in the STAR Beam
Energy Scan (BES) program.

\begin{figure}[!htb]
  \begin{center}
    \includegraphics[scale=0.4]{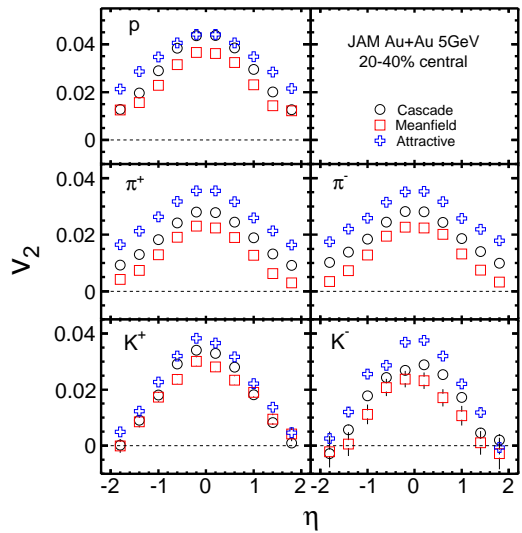}
    \caption{\label{fig:3}
   Same as Fig.~\ref{fig:2}, but
  the $v_2$ as a function of $\eta$
   in 20-40\% mid-central Au+Au collisions at $\sqrt{s_{NN}}=5$ GeV.
   }
  \end{center}
\end{figure}

We have also studied the pseudo-rapidity and transverse momentum dependence
of the $v_{2}$ in mid-central (20-40\%) Au+Au collisions.
In Fig~\ref{fig:3}, the $\eta$ dependence of $v_{2}$
for the particles ($p$, $\pi^{+}$, $K^{+}$) and corresponding antiparticles
($\pi^{-}$, $K^{-}$) are presented.
The results of the JAM model for particles and antiparticles show
a similar decreasing trend of $v_{2}$ with increase in $|\eta|$.
We observe that the values of $v_{2}$ for pions and kaons
from the attractive orbit mode
are larger than those from the cascade mode,
while $v_2$ for protons is similar to the cascade mode prediction at mid-pseudorapidity.
At the same time, from our results it is clear to see that $v_{2}$
from mean-field mode is always smaller than the results
from the cascade and attractive orbit modes for all the particles.

\begin{figure}[!htb]
  \begin{center}
    \includegraphics[scale=0.4]{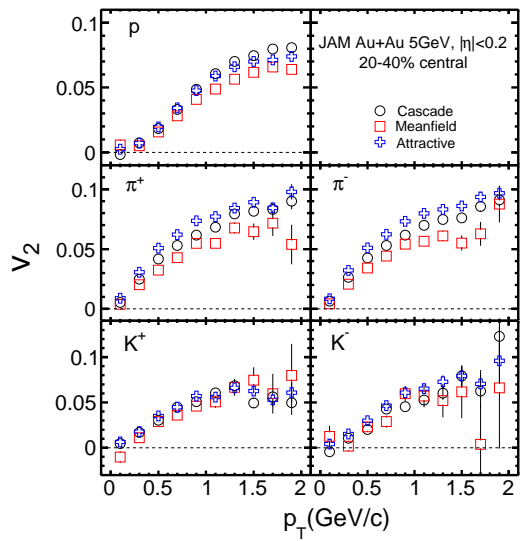}
    \caption{\label{fig:4} Same as Fig.~\ref{fig:3}, but
    the $v_2$ as a function of the transverse momentum $p_T$
 for $|\eta| < 0.2$.
   }
  \end{center}
\end{figure}

In Figure~\ref{fig:4},
we show $v_2$ for identified particles 
as a function of the transverse momentum $p_T$
for $|\eta| < 0.2$ in 20-40\% mid-central Au+Au collisions
at $\sqrt{s_{NN}}$= 5 GeV.
The results from three different modes show a similar transverse
momentum dependence in $v_2(p_T)$.
It is also observed that the proton $v_2(p_T)$
from JAM standard cascade and JAM with attractive orbit modes
are similar for low $p_T$ range.
The results of $v_2(p_T)$ from the cascade and attractive orbit mode
are larger than the result from JAM with the mean-field mode for pions.
Although the statistical error on the kaon and antikaon is relatively large,
the general increasing trend of $v_{2}(p_T)$
with increasing $p_{T}$ is still obvious. 
The difference in $v_{2}(p_T)$ between the particles
and corresponding antiparticles from JAM
is small as expected from the integrated $v_2$ results.

\begin{figure}[!htb]
  \begin{center}
    \includegraphics[scale=0.47]{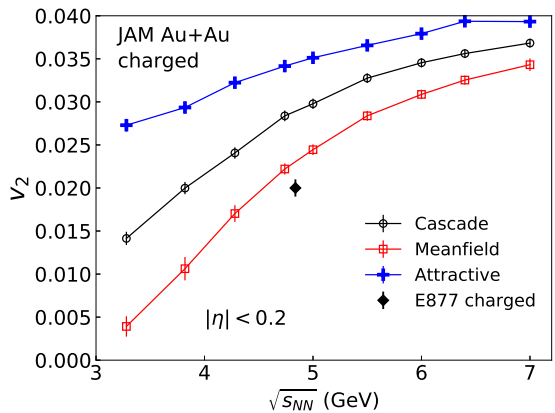}
    \caption{\label{fig:5} Beam energy dependence of
    elliptic flow $v_2$ for charged hadrons for $|\eta| < 0.2$
   from JAM in mid-central Au+Au collisions ($4.6<b<9.4$ fm).
   Data is taken from Ref.~\cite{Filimonov:2001fd}.
   }
  \end{center}
\end{figure}

Finally, in Figure~\ref{fig:5},
we compute a beam energy dependence of the elliptic flow $v_2$
for charged hadrons at mid-rapidity.
It is seen that $v_2$ from JAM attractive mode is always greater
for all beam energies up to $\sqrt{s_{NN}}=7.0$ GeV,
and the effect of mean-field is to suppress $v_2$.
We note that $v_2$ for charged hadrons above $\sqrt{s_{NN}}=7.7$ GeV
from the JAM attractive mode does not show any
enhancement relative to the JAM standard
cascade results~\cite{Nara1601}, and
the effects of hadronic mean-field on $v_2$ is very small
at SPS energies~\cite{Isse0502}.
Thus an enhancement of $v_2$ is predicted
only at the beam energy lower than 7 GeV in JAM,
which is due to the suppression of squeeze-out effect
by the softening of the EoS.
It is known that microscopic hadronic transport model predictions
including hadronic mean-field are consistent with the data
up to the top AGS energy 4.7 GeV 
~\cite{science2002,Pinkenburg:1999ya,Isse0502},
thus the scenario of the phase transition seems to be ruled out at
the beam energies less than 5.0 GeV.
However, there is no data between 5.0 and 7.7 GeV,
and it is still interesting to measure the elliptic flow by experiment
in this beam energy region in order to investigate a possible
phase transition signal of a strongly interacting matter created in
heavy ion collisions.

\section{Summary}

We have studied the effects of the hadronic mean-field
and the softening of the EoS on the elliptic flow in Au+Au collision at $\sqrt{s_{NN}}=5$ GeV within the JAM model.
The calculations of $v_{2}$ are performed within three different modes, which are cascade, mean-field, and attractive orbit, respectively. We observed that both mean-field and attractive orbit modes enhance the spectrum slope of nucleons and charged particles. 
On the other hand, we found that the value of $v_{2}$
from the attractive orbit mode is larger than the one from the cascade mode,
while the mean-field mode yields less $v_{2}$
than the results from cascade mode.
We have also presented the centrality, $p_T$ and $\eta$ dependence of $v_2$
for identified particles ($p$, $\pi^{+}$, $K^{+}$)
and corresponding antiparticles ($\pi^{-}$, $K^{-}$), respectively.
The magnitude of $v_2$ from the JAM model
for identified particles are similar to those for their antiparticles.

Our results indicate a high sensitivity of the elliptic flow on the
pressure of the system. Hadronic mean-field generates more pressure
which leads to stronger squeeze-out effect. 
On the other hand, 
the enhancement of the elliptic flow is predicted
for the attractive orbit mode which leads to
the softening of the EoS within the non-equilibrium microscopic
simulations for the first time.
The enhancement of $v_2$ is caused by a suppression of squeeze-out effects 
due to a less pressure of the system. 
Our results suggest that the enhancement of the elliptic flow in
Au+Au collision at highest baryon density region
may be used as a signal of a first-order phase transition.
For the further investigations in this direction,
a study of the EoS dependence of the elliptic flow
by the transport approach
with the EoS modified collision term~\cite{Nara:2016hbg}
may provide a useful information.

\section*{Acknowledgments}
This work is supported by the MoST of China 973-Project
No.2015CB856901, NSFC under grant No. 11575069, 11221504.
Y. N. is supported by the Grants-in-Aid for Scientific Research
from JSPS (Nos.15K05079 and 15K05098).


\end{document}